\documentclass{iopconfser}
\usepackage{subcaption}
\usepackage{makeidx}
\usepackage{gensymb} %for \degree command 
\usepackage{amsmath}
\usepackage{siunitx}

\usepackage{graphicx,indentfirst}
\usepackage{natbib,hyperref}
\setcitestyle{square,comma,numbers}
\begin{document}

\title{Spectrally shaped THz pulses from tapered dielectric waveguides}

\author{Karel Peetermans$^{1,2}$, Jonah Richards$^{3}$, Max Kellermeier$^{1}$, Klaus Floettmann$^{1}$ and Francois Lemery$^{1}$ \newline}
\affil{$^1$Deutsches Elektronen-Synchrotron DESY, Hamburg, Germany}
\affil{$^2$Department of Physics, Universität Hamburg, Hamburg, Germany}
\affil{$^3$Department of Physics and Astronomy, University of Calgary, Calgary, Canada}
\email{karel.peetermans@desy.de}
%, Wolfgang Hillert$^{1,2}$
\begin{abstract}
%In order to exploit the complete scientific potential of high-energy XFELs (electron energy $> 10\ \mathrm{GeV}$), it is necessary to provide adequate pump sources to enable pump-probe science. 
In order to exploit the complete scientific potential of user-oriented accelerator facilities, it is necessary to provide adequate pump sources to enable pump-probe science.
The users of the European XFEL have requested a THz pump source matching the X-ray repetition rate (10 Hz burst mode with up to 2700 bunches per burst) with a wide range of properties. The quest for suitable THz sources is thus a major development goal. The EuXFEL R\&D project STERN is exploring beam-based radiation generation methods using Cherenkov waveguides to satisfy these user requirements. In this work, custom waveguide geometries are designed to generate a THz pulse with arbitrary spectral content. The layer-thickness or inner radius of a wakefield structure is tapered to generate flattop or Gaussian pulses up to $1 \ \mathrm{THz}$, which can be scaled to higher frequencies.
\end{abstract}
% Beam-based radiation generation using Cherenkov waveguides is being explored by the EuXFEL R&D project STERN.
\section{Introduction}
\begin{figure}
    \centering
    \includegraphics[width=4in]{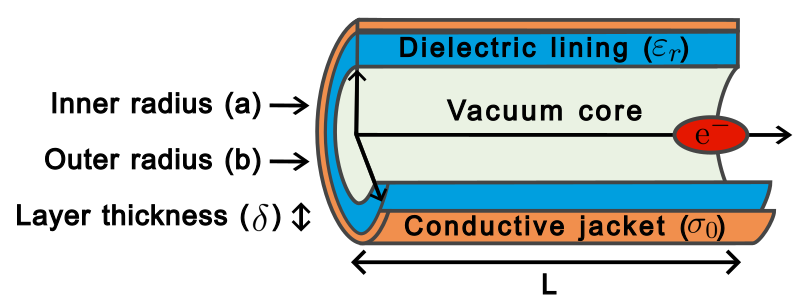}
    \caption{Sketch of a DLW and its parameters.}
    \label{fig:WaveguideSketch}
\end{figure}

Dielectric lined waveguides (DLWs) such as the one shown in Figure \ref{fig:WaveguideSketch} produce narrowband THz pulses when an electron beam traverses through~\cite{oshea2016, Floettmann2020}. 
%The pulse energy is limited by the length of the structure and its associated losses. Moreover, 
However, it has been shown that dielectrics may suffer from non-linear media responses at high field strengths, for example fused silica shows an induced conductivity near $850 \ \mathrm{MV/m}$~\cite{OSheaConductivity}. This limits the highest achievable powers and bandwidths generated from DLWs.
%have an induced conductivity with an onset near $850 \ \mathrm{MV/m}$ ~\cite{OSheaConductivity} which limits the possible powers and bandwidths generated from dielectric waveguides. 
Here we discuss the use of tapered waveguides to overcome these limitations, enabling the production of
%few-cycle (or possibly single-cycle) 
broadband pulses. With the use of a linear pulse compressor, these could be converted into intense few-cycle signals. We present a method to calculate a custom waveguide taper for a specific spectral shape of the THz pulse. This method was applied to taper either the layer-thickness or inner radius of the waveguide to produce flattop or Gaussian pulses, respectively. \par
Previous work shows tapered waveguides for broadband radiation generation from linear tapers~\cite{BANE201267}, bunch phase-space synthesis~\cite{Mayet2020}, and more recently chirp-controlled THz pulses~\cite{Liang2023}. This work focuses on balancing the amplitude at which each frequency is being produced to control the spectral distribution of the THz radiation. 
%controlling the spectral distribution of the THz radiation by balancing the amplitude at which each frequency is being produced. 
% In the following we present a method to calculate a custom waveguide taper for a specific spectral shape of the THz pulse. This method was applied to taper either the layer-thickness or inner radius of the waveguide to produce respectively flattop or Gaussian pulses.

\section{Theory}

The dispersion relation of the modes in a DLW is derived and described in~\cite{Ng_1990}. In this work, only the fundamental excited $\mathrm{TM}_{01}$ mode is taken into consideration. We therefore assume the electron beam is entering the structure on-axis. The transverse beam size at the proposed experiment location is $\sigma_r < 5 \ \mathrm{\mu m}$, justifying the model of a 1D linear charge density. \par
An estimate of the required tapering rates is found from the analytical solution that holds in the thin-layer limit. In this regime, the product of thickness $\delta$ and radius $a$ goes with the inverse of the resonant frequency $\omega$ squared. The approximate resonance condition reads~\cite{Ng_1990},
\begin{align}
    \delta (z) a (z) \approx \frac{2 \varepsilon_r c^2 }{(\varepsilon_r -1) \omega (z) ^2}.
\end{align}
Here $\varepsilon_r$ is the relative dielectric constant of the layer material.
In the case of a flattop spectrum, $\omega$ will depend linearly on the longitudinal coordinate along the waveguide $z$. A thickness or radius tapering will therefore follow an inverse quadratic curve along $z$. The density of frequencies being produced by a specific waveguide is denoted as $f(\omega)$. This function is entirely defined by the geometry of the waveguide. An infinitely long lossless waveguide with constant thickness and radius would have $f(\omega)$ resembling a delta-function at the resonant frequency. \par
When finding the density of frequencies in the resulting THz pulse, the amplitude at which a frequency is produced must also be taken into account. This value indicates how strongly the electron beam couples to the resonant mode in a given waveguide geometry. In the thin-layer approximation, is given by~\cite{Ng_1990},
\begin{align}
    &A(z) \approx \frac{4 Q}{\varepsilon_r^2 a(z) b(z)},
\end{align}
where $Q$ is the bunch charge and $b = a+\delta$ is the outer radius. If $\omega(z)$ is chosen to be bijective, $A$ can be expressed in terms of $\omega$ by substituting the inverse relation, $z=z(\omega)$. For a given frequency, the amplitude is generally stronger for a smaller waveguide diameter. In the rest of this work, the thin-layer approximation is no longer applied and the resonance condition is numerically solved.

The total frequency spectrum of a light pulse generated from one electron $E_1(\omega)$ passing through a waveguide then has the form,
\begin{align}
    E_1(\omega) = f(\omega)A(\omega).
\end{align}
%The function $f$ represents the density of frequencies being produced. The function is entirely defined by the geometry of the waveguide. 
The field produced by a complete electron bunch is the superposition of the fields generated by the individual electrons. %For typical electron bunch charges, radiation emitted incoherently can be neglected. 
The spectral distribution of the radiation emitted coherently by a charge density, $\rho_L$, is given by the longitudinal form factor~\cite{Lockmann2021},
%can be found using the longitudinal bunch form factor $F_L(\omega)$. It is the Fourier transformation of the longitudinal electron charge density $\rho_L$ given by~\cite{Lockmann2021},
\begin{align}
    F_L(\omega) = \int \rho_L (s) e^{-i s \omega /c} ds .
\end{align}
The power spectrum produced by an electron bunch is then given by,
\begin{align}
    E(\omega)&=E_1(\omega)F_{L}(\omega)\\
    &=f(\omega)A(\omega)F_{L}(\omega).
\end{align}
The method to shape the final power spectrum is based on matching $f(\omega)A(\omega)$ to the design distribution $\Tilde{f}(\omega)$. The form-factor will be taken into consideration after simulation. In the case of the EuXFEL, the longitudinal bunch form factor is typically unity over our frequency ranges of interest~\cite{Zagorodnov_2013}. However, for the sake of reducing simulation duration, longer bunches are used and the effect of the form factor is divided away in the post-processing steps. \par
THz waves travelling inside a waveguide will also experience a transport efficiency $\eta(\omega)$ from losses in the dielectric and at the metal boundary. This factor can readily be added into the equation, but will be neglected in these initial simulation results.

\section{Simulation setup}

The simulations shown below are all performed using ECHO2D~\cite{Zagorodnov2005, Zagorodnov_ECHOWebsite} to simulate the electron bunch passing through a custom geometry.  Such a geometry is calculated by first finding all the necessary radii/thicknesses to cover the frequency spectrum, as well as their corresponding amplitude $A(\omega)$. The structure parameters are sampled from the design frequency distribution $\Tilde{f}(\omega)$ divided by the amplitude so that $E(\omega)=\Tilde{f}(\omega)$.

\begin{figure}[h!]
    \centering
    \begin{subfigure}[b]{3in}
        \includegraphics[width=2.5in]{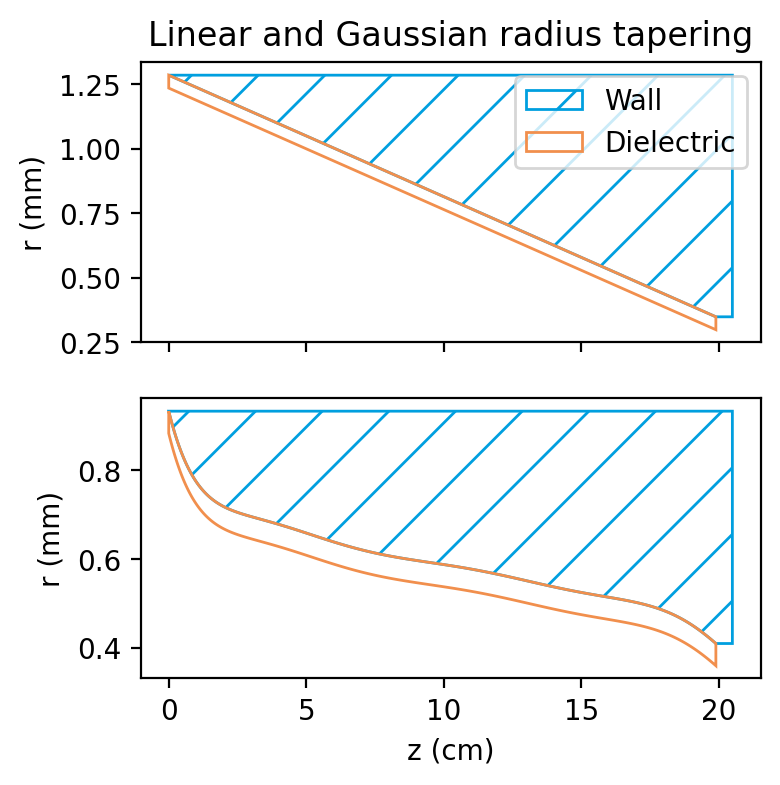}
%        \caption{Radius tapering}
    \end{subfigure}
    \hfill
    \begin{subfigure}[b]{3in}
        \includegraphics[width=2.5in]{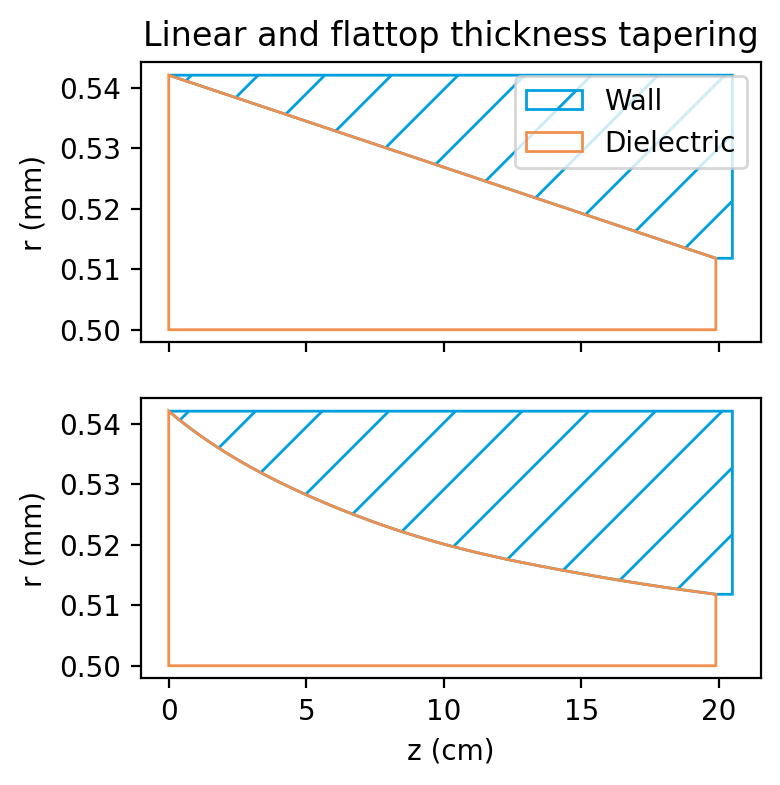}
%        \caption{Thickness tapering}
    \end{subfigure}
    \caption{Geometries used for spectral shaping from radius and thickness tapering. The top image shows a linear taper, the bottom one a custom taper to produce the desired power spectrum.}
    \label{fig:Taper-geom}
\end{figure}

The radius or thickness of the structures is chosen to be monotonically decreasing with respect to the beam direction to increase simulation speed. In experiment, such a structure might be inverted to lower losses and improve outcoupling efficiency. The effect of the beam direction on the temporal characteristics of the pulse are discussed in Section \ref{sec:temporal}. The geometries that were used in these simulations are shown in Figure \ref{fig:Taper-geom}. The simulation parameters are shown in Table \ref{tab:echo2d_params}. When the inner radius is changed, thickness is kept constant and vice versa. A third option would be to change the thickness while keeping the outer radius constant, but results are expected to be similar. The preferred tapering method depends mostly on the method of DLW production.
%(see Section \ref{sec:waveguide production}).
The electron bunch is set up as an on-axis 1D linear charge density with a Gaussian shape in the longitudinal direction defined by $\sigma_z$.

\begin{table}[h!]
    \centering
    \begin{tabular}{c|c|c}
        \hline
        Quantity & Radius Tapering & Thickness Tapering \\
        \hline
        $\sigma_z$ $[\mathrm{\mu m}]$ & $50$ & $25$ \\
        $z_{\text{step}}=y_{\text{step}}$ $[\mathrm{\mu m}]$& $10$ & $10$ \\
        Mesh Length $[\mathrm{cm}]$ & 20 & 20 \\
        Radius $[\mathrm{\mu m}]$ & $349 - 1286$ & $500$ \\
        Thickness $[\mathrm{\mu m}]$ & $50$ & $11 - 43$ \\
        Length $[\mathrm{c m}]$ & $20$ & $20$ \\
        Rel. dielectric const. & $3.81$ & $3.81$ \\
        \hline
    \end{tabular}
    \caption{ECHO2D simulation parameters for radius and thickness tapering}
    \label{tab:echo2d_params}
\end{table}

In the post-processing step, instantaneous power and energy are found by integrating the Poynting-vector~\cite{BANE201267},
\begin{align}
S(t) = 2 \pi \int_0^b E_r(r,t) H_{\phi} (r,t) r dr.
\end{align}
The integral is numerically evaluated over a field-monitor spanning from the axis to the edge of the structure. Since no Ohmic and dielectric losses are being considered, the energy in the THz pulse equals the energy lost by the electron bunch up to numerical accuracy.

\section{Simulation results}
\subsection{Power spectra}
\begin{figure}[h!]
    \centering
    \begin{subfigure}[b]{3in}
        \centering
        \includegraphics[width=2.5in]{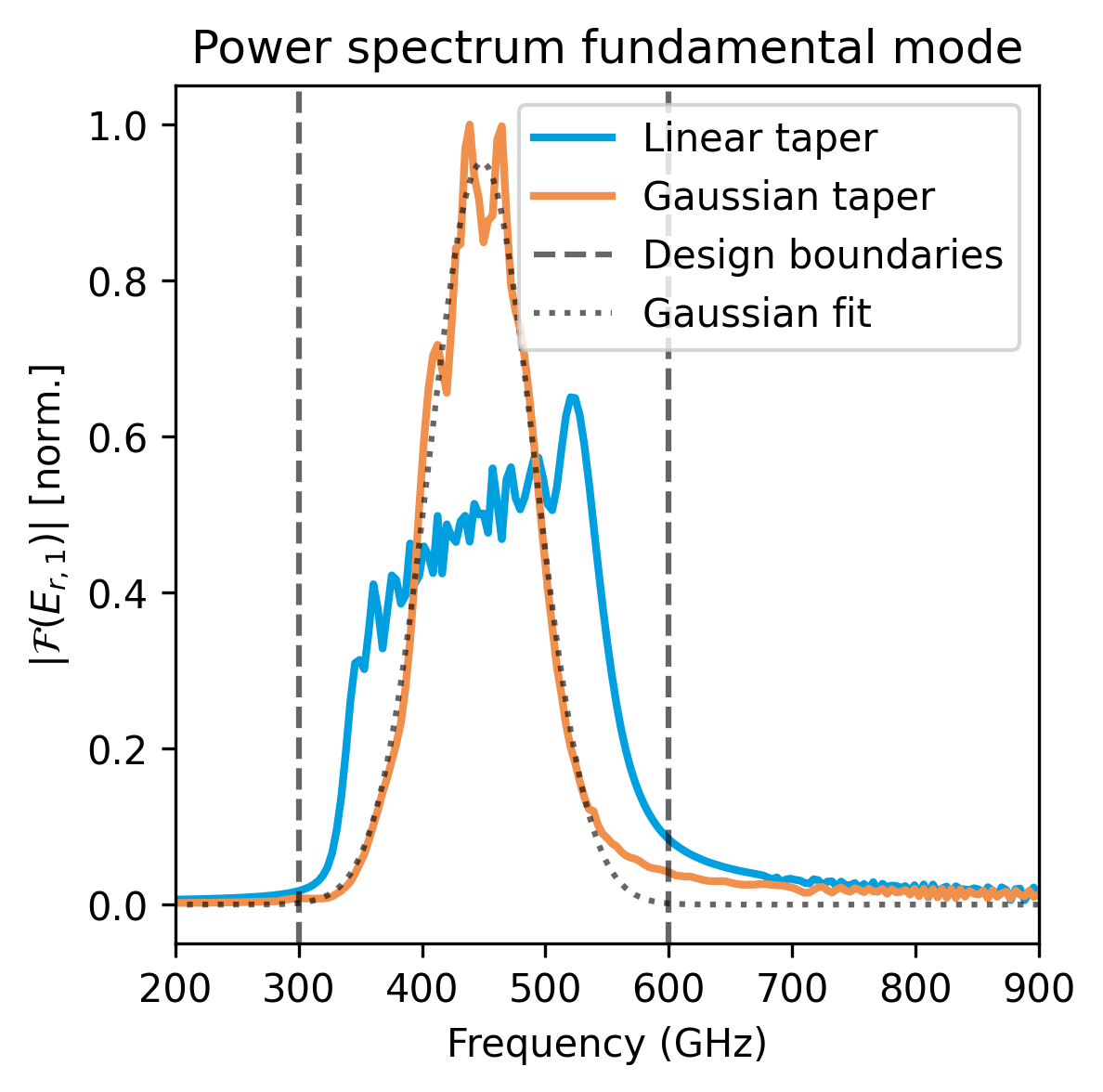}
        \caption{Radius tapering}
    \end{subfigure}
    \hfill
    \begin{subfigure}[b]{3in}
        \centering
        \includegraphics[width=2.5in]{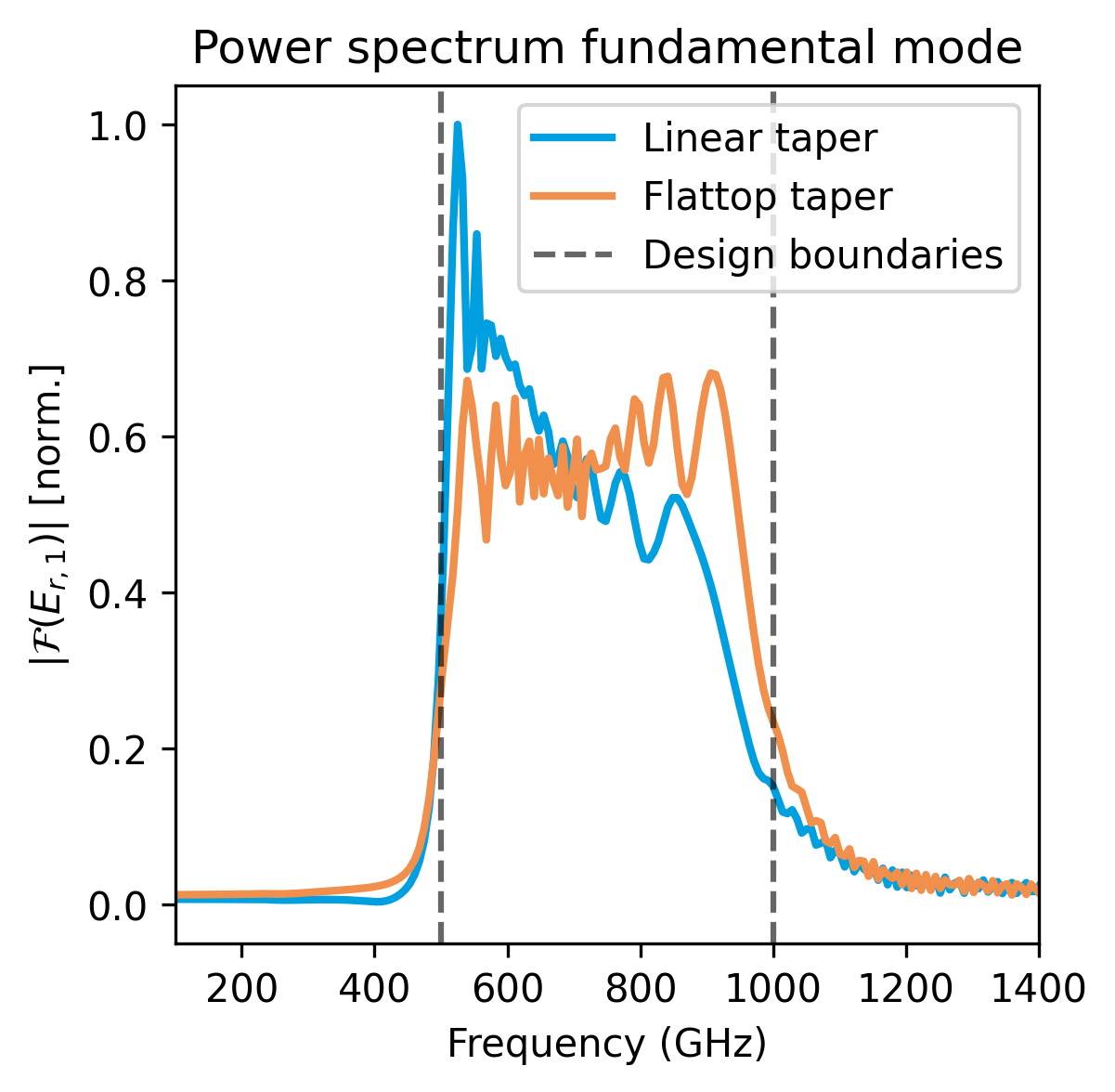}
        \caption{Thickness tapering}
    \end{subfigure}
    \caption{Comparison of the generated power-spectra. Both plots show the comparison of the custom tapering with a linear taper covering the design boundaries.}
    \label{fig:PowerSpectrumComparison}
\end{figure}

%The geometries and parameters shown above were implemented and their wakefields simulated on a Windows machine. 
The geometeries and parameters shown above were implemented and their wakefields simulated using ECHO2D version 3.5.
%Simulations of this resolution take up to a few hours to complete. 
Due to the high resolution, the computation time spans from 1 to 5 hours. The THz spectrum is virtually measured with an electric field monitor at the structure exit at $r=300\ \mathrm{\mu m }$. The results are shown in Figure \ref{fig:PowerSpectrumComparison}. The Gaussian fit returned a FWHM of $99.2 \ \mathrm{GHz}$, compared to the design value of $75 \ \mathrm{GHz}$. The approximate shape of the desired spectrum is being reproduced very well with a relatively short structure. For larger bandwidths, the need for tailored tapers becomes evident, as linear tapers tend to over-produce one side of the spectrum. The general lack of smoothness in the spectra is expected to reduce when increasing numerical accuracy and waveguide length. 

\subsection{Temporal frequency chirp and pulse compression}\label{sec:temporal}

\begin{figure}[h!]
    \centering
    \begin{subfigure}[b]{3in}
        \centering
        \includegraphics[width=2.5in]{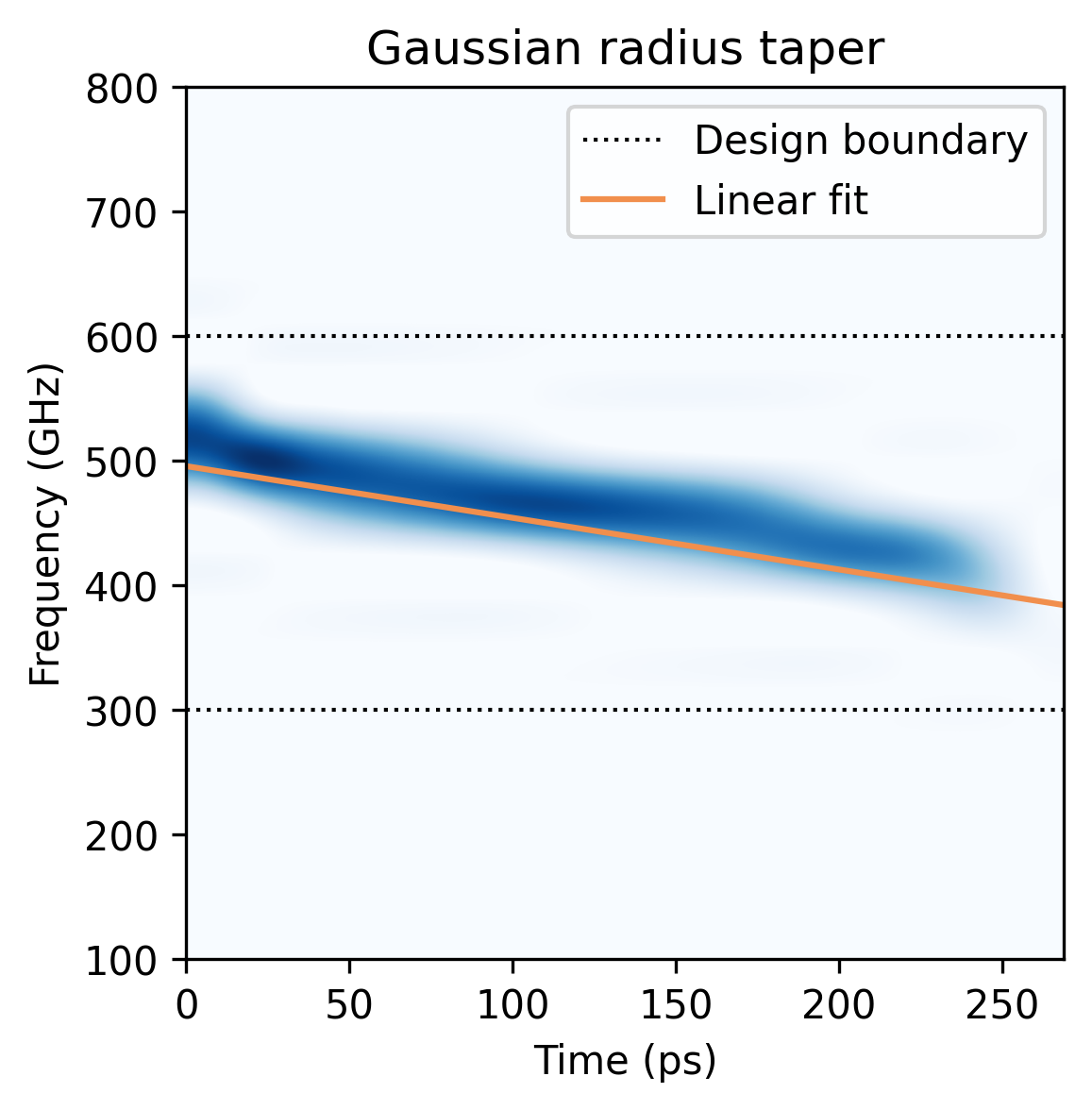}
        \caption{Radius tapering}
    \end{subfigure}
    \hfill
    \begin{subfigure}[b]{3in}
        \centering
        \includegraphics[width=2.5in]{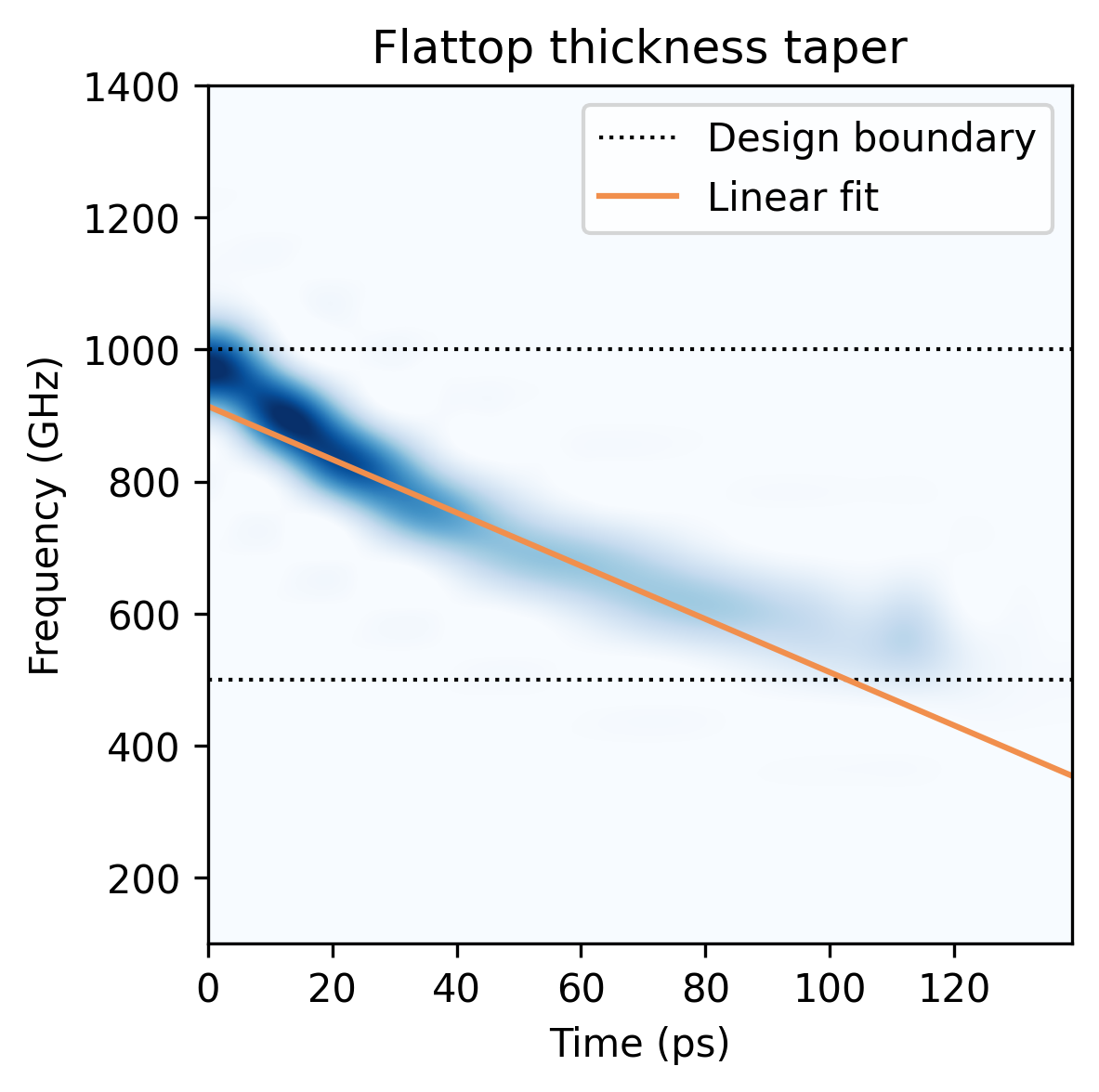}
        \caption{Thickness tapering}
    \end{subfigure}
    \caption{Comparison of the generated frequency chirps. The spectrograms were smoothed using the Lanczos algorithm.}
    \label{fig:SpecGrams}
\end{figure}

A THz pulse with bandwidths of around an octave (i.e. a spectrum from $f_i$ to $2 f_i$) can be compressed using a linear compressor~\cite{Lockmann2021}. Such a compressor could be built using gratings, prisms or even photonic crystals. There are limited material options with high transmittance in the THz regime, which is why we believe grating-based compressors would achieve the best performance.
%In the waveguide design no attention was paid to the temporal characteristics of the frequency chirp. The performance of such a compressor is limited in this case, but was nonetheless applied numerically. 
To emulate the effect of the compressor stage, a linear function was fitted to the spectrogram and a compression with the resulting parameters was applied. These spectrograms are shown in Figure \ref{fig:SpecGrams} together with their linear fit. In both cases the frequency decreases monotonically with time. Because of dispersion, the waveguide naturally compresses the pulse. If the geometry was flipped, the chirp would have a rising slope and a longer pulse length. The choice between the two can be made based on the user requirements and the available compressor. \par
\begin{figure}[]
    \centering
    \includegraphics[width=5in]{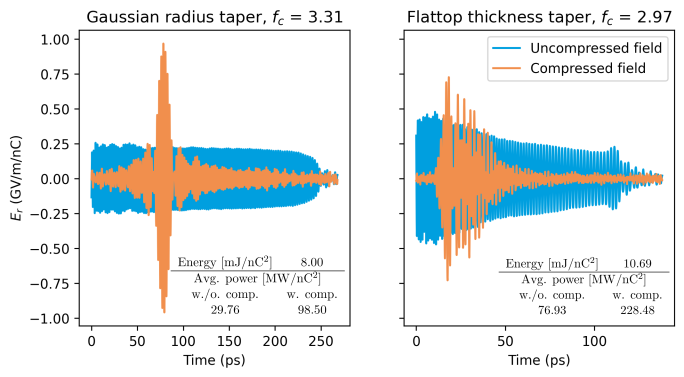}
    \caption{Comparison of the electric field at the end of the structure.}
    \label{fig:FieldComp}
\end{figure}

The factor of compression, defined as the fraction of the RMS pulse lengths, is around $3$ in both cases. The compressed electric fields reach up to $1 \ \mathrm{GV/m/nC}$ with pulse powers up to $228 \ \mathrm{MW/nC^2}$. These simulation results are independent of total bunch charge and can readily be rescaled to standard units using the charge of the bunch. The fields before and after compression are shown in Figure \ref{fig:FieldComp}. In the waveguide design no explicit attention was paid to the temporal characteristics of the frequency chirp. The performance of a linear compressor could therefore be further investigated and improved. 


%\section{Waveguide production}\label{sec:waveguide production}

%The geometries shown above are sampled down to $500$ discrete slices. Each slice is therefore $400 \mathrm{\mu m}$ in length. At frequencies $< 1 \ \mathrm{THz}$, layer thicknesses exceed $10 \ \mathrm{\mu m}$. Therefore, a 3D printer with low-loss plastic or even fused silica~\cite{Kotz2017} could be used to produce the dielectric structure. As an example, the ASIGA MAX X27~\cite{ASIGA} has a pixel size of only $27 \mathrm{\mu m}$.  Afterwards such a structure could be coated with a conductive metal (e.g. copper).\par Alternatively, a 3D printer could be used to precisely print the metal outside of the waveguide after which the inside is coated with a dielectric. Such coatings can be applied with atomic layer deposition. Producing custom power spectra from tapered DLWs can therefore be achieved with currently available technology.

\section{Conclusions and outlook}
A method to design tailor-made waveguide geometries was presented. 
%Structures with tapered inner radius or layer thickness are generated to produce an arbitrary power spectrum in the THz regime.  \par
This work shows the design of bandwidths spanning an octave in the shape of a flattop and a Gaussian curve. The structures were implemented into ECHO2D and the resulting spectra agree with the intended design. The simulations can readily be expanded to encompass other spectral shapes or frequency regions. \par

The EuXFEL user community has requested pulse bandwidths of $5\%$ to $100\%$ over the frequency range of $0.3-30 \mathrm{THz}$~\cite{Zalden_2018}. Hence, an array of waveguides could be produced to cover the most commonly requested spectra.
%Hence, a survey over interested THz users could be held to produce an array of waveguides covering commonly requested spectra. 
%~\cite{Kotz2017}
These structures could feasibly be created using a 3D-printed low-loss plastic or fused silica capillary~\cite{Kotz2017}, coated on the outside with a conductive metal. Alternatively, the outer metal of the waveguide could be printed after which the inside could be covered with atomic layer deposition. 
During operation, the user could select the waveguide to send the electron bunch through and produce the preferred pump/probe pulse. \par
To make this study more realistic, losses should be taken into account when designing the waveguide. This will likely require an iterative generation process similar to the one featured in~\cite{Liang2023}. Such an implementation could then also provide control over the temporal shape of the spectrum to achieve better pulse compression. This way, even single-cycle pulses are expected to be achievable. Another outstanding challenge is the production and testing of these structures with an existing THz source or an electron beam to compare the simulations with experiments. 

\section*{Acknowledgements}
This work was supported by funds of the European XFEL R\&D program. The authors would like to thank Igor Zagorodnov for his support in the use of ECHO2D. 
\newpage
%\printbibliography
%\label{bibliography}
%\section*{References}
\bibliographystyle{iopart-num-mod}
\bibliography{bib}
\end{document}